 \documentclass[final,1p,times]{elsarticle}
\usepackage{color}

\usepackage{amssymb}

\begin{document}

\begin{frontmatter}



\title{Quantum correlation dynamics in photosynthetic processes assisted by molecular vibrations}

\author{G. L. Giorgi}
\address{INRIM, Strada delle Cacce 91, I-10135 Torino, Italy }
\ead{g.giorgi@inrim.it}
\cortext[cor1]{Corresponding author}

\author{M. Roncaglia}
\address{INRIM, Strada delle Cacce 91, I-10135 Torino, Italy }

\author{F. A. Raffa}
\address{Politecnico di Torino, Dipartimento di Scienza Applicata e Tecnologia,\\
Corso Duca degli Abruzzi 24, I-10129 Torino, Italy}

\author{M. Genovese}
\address{INRIM, Strada delle Cacce 91, I-10135 Torino, Italy }

\begin{abstract}
During the long course of evolution, nature has learnt how to exploit quantum effects. 
In fact, recent experiments reveal the existence of quantum processes whose coherence extends over unexpectedly long time and space ranges.  
In particular, photosynthetic processes in light-harvesting complexes display a typical oscillatory dynamics ascribed to quantum coherence. 
Here, we consider the simple model where a dimer made of two chromophores is strongly coupled with a quasi-resonant vibrational mode. 
We observe the occurrence of wide oscillations of genuine quantum correlations, between electronic excitations and the environment, represented by vibrational bosonic modes. Such a quantum dynamics has been unveiled through the calculation of the negativity of entanglement and the discord, indicators widely used in quantum information for quantifying the resources needed to realize quantum technologies. We also discuss the possibility of approximating additional weakly-coupled off-resonant vibrational modes, simulating the disturbances induced by the rest of the environment, by a single vibrational mode.
 Within this approximation, one can show that the off-resonant bath behaves like a classical source of noise.
\end{abstract}


\end{frontmatter}



\section{Introduction}

The existence of coherence, caused by the interference of probability amplitude terms, is one of the distinctive traits of quantum mechanics.  Oscillatory behaviours, ubiquitously observed in quantum systems, are  the consequence of such coherent phenomena. 
The fact that light-harvesting complexes were experimentally proven to exhibit oscillatory electronic dynamics has stimulated a deep debate about the nature of such oscillations, and, consequently, the possible role played by quantum mechanics in biologically functional systems \cite{Engel07, Calhoun09,Panitchayangkoon10,Collini10, Harel12}. Clearly, oscillations can also be found in completely classical systems (the appearance of oscillatory electronic dynamics within classical models was studied in  \cite{Miller12, Briggs11}).
Thus, it is of the utmost importance to give a precise characterization of these oscillations by means of some quantumness quantifier, in order to unveil the basic mechanisms adopted by nature, which in future could possibly inspire new energy technologies.                                                                                                                                                                                                                                                                                                                                                                                                                                                                                                                                                                                                                                                                                                                                                                  
%

The question of whether  coherence in light-harvesting complexes has a quantum or classical origin was addressed in a series of recent seminal works \cite{Olaya08,Thorwart09,Caruso09,Wilde10,Sarovar10,Caruso10,Fassioli10, Whaley11}. In Ref. \cite{Wilde10}, Wilde \textit{et al.} faced the problem from the macro-realism point of view using the  Leggett-Garg inequalities to test whether the system dynamics is compatible with classical theories, while in Refs. \cite{Thorwart09,Sarovar10,Caruso10,Fassioli10, Whaley11,Gauger11,zhu12} the presence of entanglement among the electronic degrees of freedom was used to assess the genuine quantum character of the whole system.

As living objects are embedded in their own environment, the dynamical behaviour of electronic excitations is necessarily influenced by the presence of other degrees of freedom, mainly phonons, whose coupling to the system has been identified to be one of the possible causes of efficient transport \cite{Renger01,mohseni08, Plenio08, Rebentrost09,wu10,delrey}. 
 Stimulated by this observation, in Ref. \cite{oreilly14}, O'Reilly and Olaya-Castro recently investigated non-classical features of the molecular motions and phonon environments in a prototype dimer that can be found in light-harvesting antennae of cyanobacteria \cite{Womick11}, cryptophyte algae \cite{Doust04,Novoderezhkin10} and higher plants \cite{Liu04, Barros09, Novoderezhkin04b}. A characterization of quantumness was performed by means of the Mandel $Q$-parameter and the Glauber-Sudarshan quasi-probability $P$ distribution, whose negative regions in phase space are not compatible with any classical description of the coupled dynamics.

Nevertheless, albeit $Q$-parameter and $P$-function negativity represent a significant way of quantifying the quantumness of a system, they do not catch thoroughly the ultimate quantumness represented by non-classicality of correlations. For instance, the $P$-function of an Einstein-Podolski-Rosen(EPR) pair is positive, in spite of the deep quantum nature of this state \cite{genovesepr}.

In this paper, we want to overcome this drawback by directly quantifying the degree of quantumness developed during the coupled exciton-vibration dynamics through the use of entanglement and of quantum discord between system and environment.
On the one hand, entanglement is usually viewed as an utterly fragile property at room temperatures, as it can be easily destroyed by decoherence \cite{entanglement}. Despite these caveats, however, its presence was predicted in different biological processes  \cite{Olaya08,Sarovar10,Caruso10,Fassioli10, Whaley11,Thorwart09,Gauger11,zhu12}. 
On the other hand,  there can exist quantum correlations also in the absence of entanglement, as witnessed by many  quantum protocols. These quantum correlations are captured by quantum discord, whose definition originates from two definitions of the classical mutual information whose quantized versions turn out to be nonequivalent \cite{discord}. Quantum discord provides us a criterion of quantumness that is both necessary and sufficient, in contrast, for instance, with the use of the Mandel parameter or the $P$-distribution. In the case of entanglement, we will make use of the negativity  \cite{negativity}, which is a sufficient criterion itself, and of a lower bound for the entanglement of formation \cite{albeverio}. 
The analysis will be performed by considering the dimer-exciton system both in the presence and in the absence of decoherence effects induced by low-energy modes in the phonon environment. It will be also interesting to monitor the quantum character of the bath by quantifying the entanglement between the bath itself and the system.

\section{Results}

\subsection{Model}
In Ref.~\cite{oreilly14}, it was shown  that in prototype dimers present in a variety of biological systems, efficient
vibration-assisted energy transfer in the sub-picosecond timescale and at room temperature
can appear. It was also shown that non-classical fluctuations of collective pigment motions are dynamically created. Based on these observations, it was suggested that a connection may exist between these fluctuations and the high efficiency of the process. The model employed consists of an effective dimer coupled to an undamped bosonic mode. Despite its simplicity, it captures the essential features of the problem.

Let us just briefly recall the physical model. 
A dimer is composed of two chromophores whose Hamiltonian reads 
\begin{equation}\label{HEL}
H_{\rm el} = \sum_{i=1,2}\varepsilon_i \sigma^+_i \sigma^-_i + V(\sigma^+_1\sigma^-_2+\sigma^+_2\sigma^-_1), 
\end{equation}
where $\varepsilon _i$ is the energy of the excited level of the $i$-th chromophore and $V$ is the inter-chromophore coupling. The operators $\sigma^+_i$ ($\sigma^-_i$) create (annihilate) an electronic excitation at site $i$ and are expressed in terms of Pauli matrices, $\sigma^\pm_i$ $=$ $(\sigma_x \pm i \sigma_y)/2$. Note that $\sigma^+_1 \sigma^-_1$ and $\sigma^+_2 \sigma^-_2$ are occupation number operators with eigenvalues 0, 1, while $\hat{N}$ $\doteq$ $\sigma^+_1 \sigma^-_1 + \sigma^+_2 \sigma^-_2$ is a constant of motion of $H_{\rm el}$ with eigenvalues 0, 1, 2. The dimer is strongly coupled to a quantized vibrational mode of frequency $\omega_\textrm{vib}$, the phonon Hamiltonian being
\begin{equation}\label{HVIB}
H_{\textrm{vib}}= ~\omega_{\textrm{vib}}(b_1^\dag b_1+ ~b_2^\dag b_2),
\end{equation}
where $b_j^\dag$ ($b_j$), $j = 1, 2$, are bosonic operators which create (annihilate) one phonon of the vibrational mode of the $i$-th chromophore, so that one can define the corresponding number operators, $\hat{n}_j \doteq b_j^\dag b_j$. Finally, the electronic excited states interact with their local vibrational environments with strength $g$. For the corresponding interaction Hamiltonian one has
\begin{equation}\label{HELVIB}
H_{\textrm{el-vib}} = g \sum_{i=1,2}\sigma^+_i\sigma^-_i(b^\dag_i+ b_i) .
\end{equation}
Combining  Eqs. (\ref{HEL}-\ref{HELVIB}) and restricting the dynamics to one electronic excitation, one obtains the effective Hamiltonian
\begin{equation}
 H_{\textrm{ex-vib}} = \frac{\varepsilon_1-\varepsilon_2}{2} \sigma_z +  V \sigma_x - \frac{g}{\sqrt 2} \sigma_z (b_{-}^{\dag}+b_{-}) +\omega_{\textrm{vib}} b^{\dag}_{-} b_{-}, \label{hsys}
\end{equation}
where $\sigma_z$ $=$ $\sigma^+_2\sigma^-_2-\sigma^+_1\sigma^-_1$ and $\sigma_x$ $=$ $\sigma^+_1\sigma^-_2+\sigma^+_2\sigma^-_1$ and where $b_{-}^{\dag}=(b^\dag_1 - b^\dag_2)/\sqrt{2}$ is the relative  displacement phonon mode. A detailed derivation of Eq. (\ref{hsys}) is given in the appendix.

\subsection{Exciton-vibration correlations}
The quantum character of the bosonic field was analyzed by the authors of Ref.~\cite{oreilly14} using both 
the Glauber-Sudarshan $P(\alpha)$-function and the Mandel factor $Q=(\langle \hat{n}^2 \rangle-\langle \hat{n} \rangle^2)/\hat{n}-1$.
Both $P(\alpha)<0$ and $Q<0$ are used as a sufficient criterion to verify the presence of quantumness in the system. For instance, a negative value of $P(\alpha)$ implies that the density matrix cannot be expressed as a statistical mixture of coherent states and, then,  does not admit a classical interpretation.  Taking the numerical parameters from the cryptophyte antennae phycoerythrin (PE545) it was shown that both $Q(t)$ and the $P(\alpha)$-function exhibit partial non-classical behaviour, that is, there are regions of time for $Q(t)$ and regions of $\alpha$ for $P$ where quantum fluctuations are detected. These regions would vanish in the absence of dipole coupling ($V=0$).

A different approach to nonclassicality has emerged in the literature, which focusses on information-theoretic aspects of correlations and  has been shown to be deeply inequivalent to the quantum phase space and quasi-probability distribution criteria \cite{ferraro12}. Thus, the concepts of entanglement and quantum discord, considered as fundamental resources in many quantum information protocols, offer an independent and conceptually stronger alternative of verifying the quantum character of the process under study.
 
%
%

In order to detect the entanglement between the dimer and the bosonic mode we will resort to the Peres-Horodecki criterion, which states that if the density matrix fails to be positive under partial transposition \cite{entanglement}, then we are necessarily in the presence of entanglement. As a quantifier, we use the negativity of entanglement $E_N$, which amounts to the sum of the negative eigenvalues of the dynamical density matrix after partial transposition. 
Indeed, any bipartite state has the form $\varrho = \sum_{ijkl} c_{ijkl} |i\rangle \langle j | \otimes |k\rangle \langle l| $ and the partial transpose map $I \otimes T (\varrho)$  transforms  it into  $\varrho = \sum_{ijkl} c_{ijkl} |i\rangle \langle j | \otimes |l\rangle \langle k| $. As any bipartite separable state can be written, by definition of separability, as $\varrho=\sum_i p_i\varrho_A^{(i)}\otimes\varrho_B^{(i)}$, $I \otimes T (\rho)$ would map it into $\varrho^{T_B}=\sum_i p_i\varrho_A^{(i)}\otimes(\varrho_B^{(i)})^T$, which is a perfectly acceptable density matrix. Then, all the eigenvalues $\lambda_i$ of $\varrho^{T_B}$ are real, positive, and obey $\sum_i\lambda_i=1$. If, on the other hand, some of the eigenvalues of $\varrho^{T_B}$ are negative,  we can conclude that $\varrho$ does not admit a factorized form.
Based on these considerations, the quantity we are going to calculate is an entanglement witness, even though  it is not  a proper measure, aside from some special cases (a recent example can be found in Ref. \cite{roncaglia}).  States that are entangled even though their negativity vanishes are known as bound entangled states, their main characteristic being that it is not possible to obtain pure entangled states from them by means of local operations and classical communication.

 The negativity is also useful to calculate a lower bound for the entanglement of formation of a bipartite state \cite{albeverio}. Indeed, entanglement of formation represents one of the most meaningful measures of entanglement,   as its regularized version quantifies the minimal cost needed to prepare quantum states in terms of EPR pairs.

Together with  entanglement, it is also of  interest to monitor the behaviour of quantum discord ${\cal D}$, which is a more general definition of quantumness with respect to entanglement, being nonzero even in case of factorized states whose correlations do not admit any classical interpretation \cite{discord}. The definition of discord is given in the appendix. Here we anticipate that, given two parties $A$ and $B$ the quantification of the correlations between them is the goal of our study, it measures the minimum amount of disturbance introduced in the state of party $A$ because of a measurement process performed on party $B$. 

\begin{figure}
\begin{center}
\includegraphics[width=7cm]{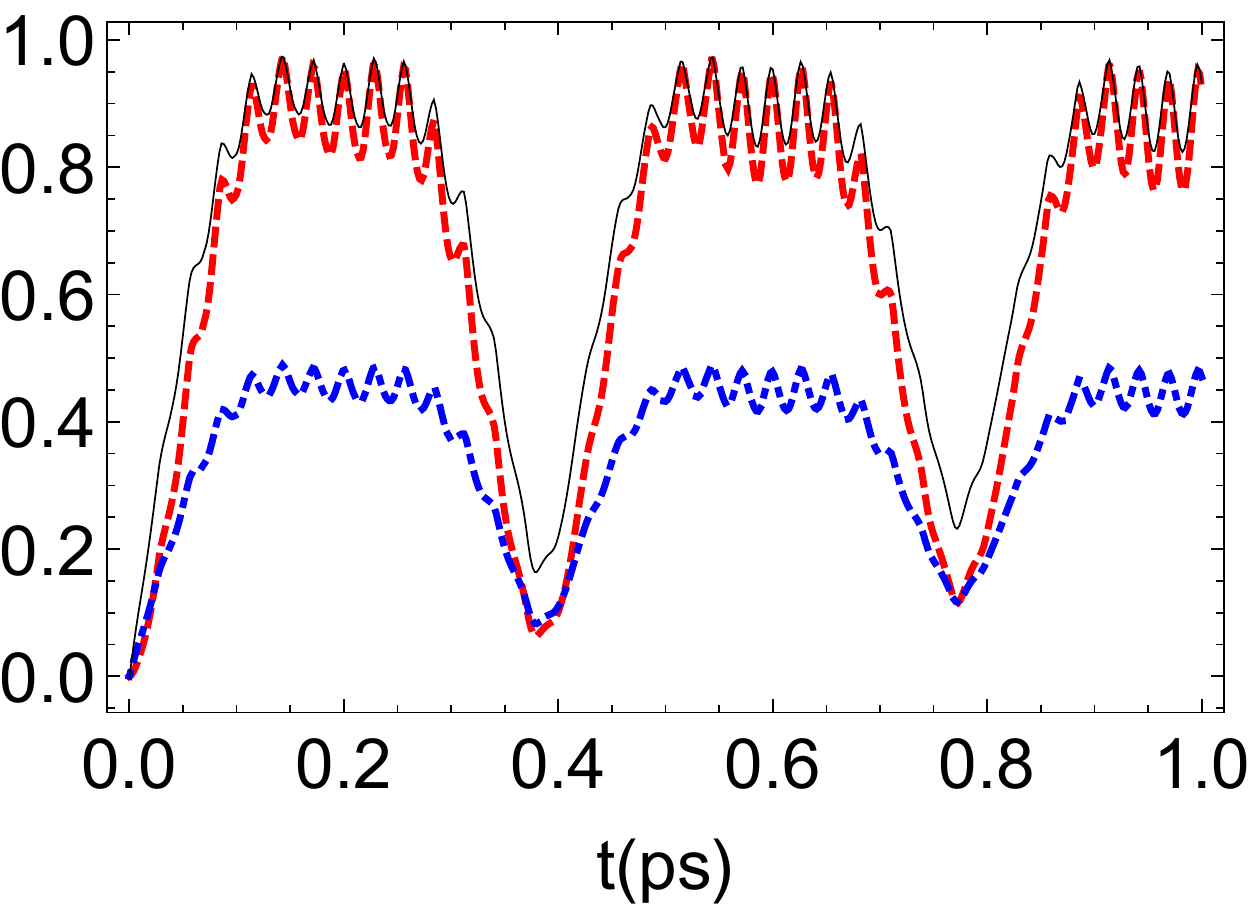}
\caption{Quantum discord (red, dashed), entanglement negativity (blue, dotdashed) and entanglement of formation lower bound (black) as a function of time.  The parameters used, taken from Refs. \cite{Doust04,Novoderezhkin10} are the following: $ \varepsilon_1- \varepsilon_2=1042~\textrm{cm}^{-1}, \; V=92~ \textrm{cm}^{-1},\; \omega_{\textrm{vib}}=1111~\textrm{cm}^{-1}$, and $g=267.1~ \textrm{cm}^{-1}$.}
\label{fig1}
\end{center}
\end{figure}

The behaviour of these two quantities is shown in Fig. \ref{fig1} as a function of time. The system is prepared at $t=0$ in the Gibbs (thermal) state at room temperature $T=270~K$ of one of the vibrational modes, which is initially uncorrelated with the 
dimer state $|X_+ \rangle$:
\begin{equation}
\rho(0) =|X_+ \rangle\langle X_+|\otimes\varrho^\textrm{th}_{\rm vib}.
\end{equation}
Here, $H_{\rm el} |X_\pm\rangle$ $=$ $\lambda_\pm |X_\pm\rangle$, where $\lambda_{\pm}=\pm\sqrt{( \varepsilon_1-\varepsilon_2)^2+4V^2}/2$.
We assume that the frequency of the mode is  much larger than the thermal energy scale $\omega_{\textrm{vib}}\gg K_\textrm{B}T$. The time evolution is calculated by solving the Liouville-von Neumann equation $\dot{\rho}=-i[H_{\textrm{ex-vib}} ,\rho]$. 
In principle, as we deal with an infinite-dimensional system the eigenstates of which are not Gaussian functions, we would need to calculate an infinite number of matrix elements in order to determine the exact full dynamics of the state and its correlations. However, taking into account that, in the regime $\omega_{{\rm vib}}\gg K_B T$, at the initial time only a few matrix elements are significantly populated, a truncation in the number of excitons is a very good approximation. The truncation scheme consists of neglecting all the matrix elements between the threshold $\tilde{n}$ and  $\tilde{n}+1$. In the model under study 
 we can safely take $\tilde{n}=5$, as the results do not change (within the machine error) for higher thresholds. 

Negativity of entanglement is calculated by applying transposition to the dimer part of the density matrix while leaving the bosonic mode unchanged. In the case of quantum discord, the dimer represents the part under measurement. In principle, an optimization over a complete  positive operator valued measure (POVM) with elements  $\{E_j^B\}$ should be performed in order to get the optimal measurement. However, for the sake of simplicity, we will limit ourselves to the class of orthogonal projectors. In all the cases known in literature, orthogonal projectors give a very tight bound and can be safely used without any appreciable qualitative change.

Working in the $\omega_{{\rm vib}}\gg K_B T$ regime implies that the dynamics is largely dominated by the coherent oscillation between $|X_+,0\rangle$ and $|X_-,1\rangle$, which absorbs most of the spectral weight of the whole density matrix. We then expect, out of the exact dynamics, a kind of ``two-qubit'' Rabi-like oscillation that unavoidably generates entanglement. 
As it can be observed, the quantumness witnessed by the presence of negativity and quantified by bipartite discord and entanglement of formation, is actually present for any $t>0$, even in the time windows when the  $Q$-parameter and the $P$-distribution fail to detect it, that is, when the Rabi-like oscillations reach their minimum \cite{oreilly14}. Around those regions, the persistence of quantum effects clearly indicates that, beyond the main oscillation, the multi-mode character of the vibration plays an important role. Interestingly, Fig. \ref{fig1} shows that all quantifiers are close to the values corresponding to maximally entangled states for most of the time interval analyzed (cf. the value $1/2$ for the negativity). We also emphasize that, since entanglement is successfully detected by the negativity for any $t$, the problem of determining the possible occurrence of bound entanglement is removed \cite{entanglement}.

%

\subsection{Single-mode description of thermal noise}

So far, we have discussed the interaction of the dimer with a single high-energy vibronic mode. In general,  low-frequency phononic modes are also present and   cannot be neglected, as their existence, albeit taken into account at a higher perturbation order, would cause decoherence and dissipation of the system under investigation.

A phonon bath is described by a collection of independent harmonic oscillators 
that can cause incoherent transitions between the system eigenstates. The position of each phonon is indeed coupled to the exciton operator $\sigma_z$, while it is decoupled from the vibrational mode.
The bath Hamiltonian can be written as
\begin{equation}
H_B=\sum_k \omega_k b_k^\dag b_k ,
\end{equation}
while the system-bath interaction Hamiltonian takes the form 
\begin{equation}
H_I=\sum_k g_k (\sigma_z\otimes  1\!\!1_{\rm{vib}}) (b_k^\dag+b_k).
\end{equation}

In principle, a hierarchical expansion of the interaction could  be employed to obtain the reduced system dynamics \cite{zhu11}.
The second order (Bloch-Redfield) perturbation theory would greatly simplify the calculation. However, due to the strong detuning between the system and the bath, non-Markovian effects, which would not be captured by that method, are expected to be relevant. 
 Remarkably, a qualitative description of the phenomenon can be obtained by drastically simplifying the approach. In the absence of the environment, as illustrated in the previous section, the dynamics is deeply influenced by the approximate degeneracy of the levels  $|X_+ , 0\rangle$ and $|X_-, 1\rangle$. As the initial population of $|X_+ ,0\rangle$ at room temperature is close to $1$, coherent oscillations between those two levels are observed which, among other effects, also determine the establishment of the quantum correlations described in Fig. \ref{fig1}. As the eigenmodes of the bath lie in a region of the energy spectrum that is far apart from the frequencies of the closed system, it is natural to ask whether and to what extent the internal structure of the environment matters.  To this end, we employ a minimal approach replacing the whole environment with a single, low-energy bosonic mode $k_0$. Since, because of the approximation method used, we are not in the presence of a true bath, as  in any few-body problem, a continuous flow of information (which is the cause of oscillations) between the system and the low-frequency mode is expected to take place instead of the relaxation behaviour typical of decoherence. Notice that in this way the  non-Markovian character of the evolution (even if in an approximate form) is kept.

\begin{figure}
\begin{center}
\includegraphics[width=7cm]{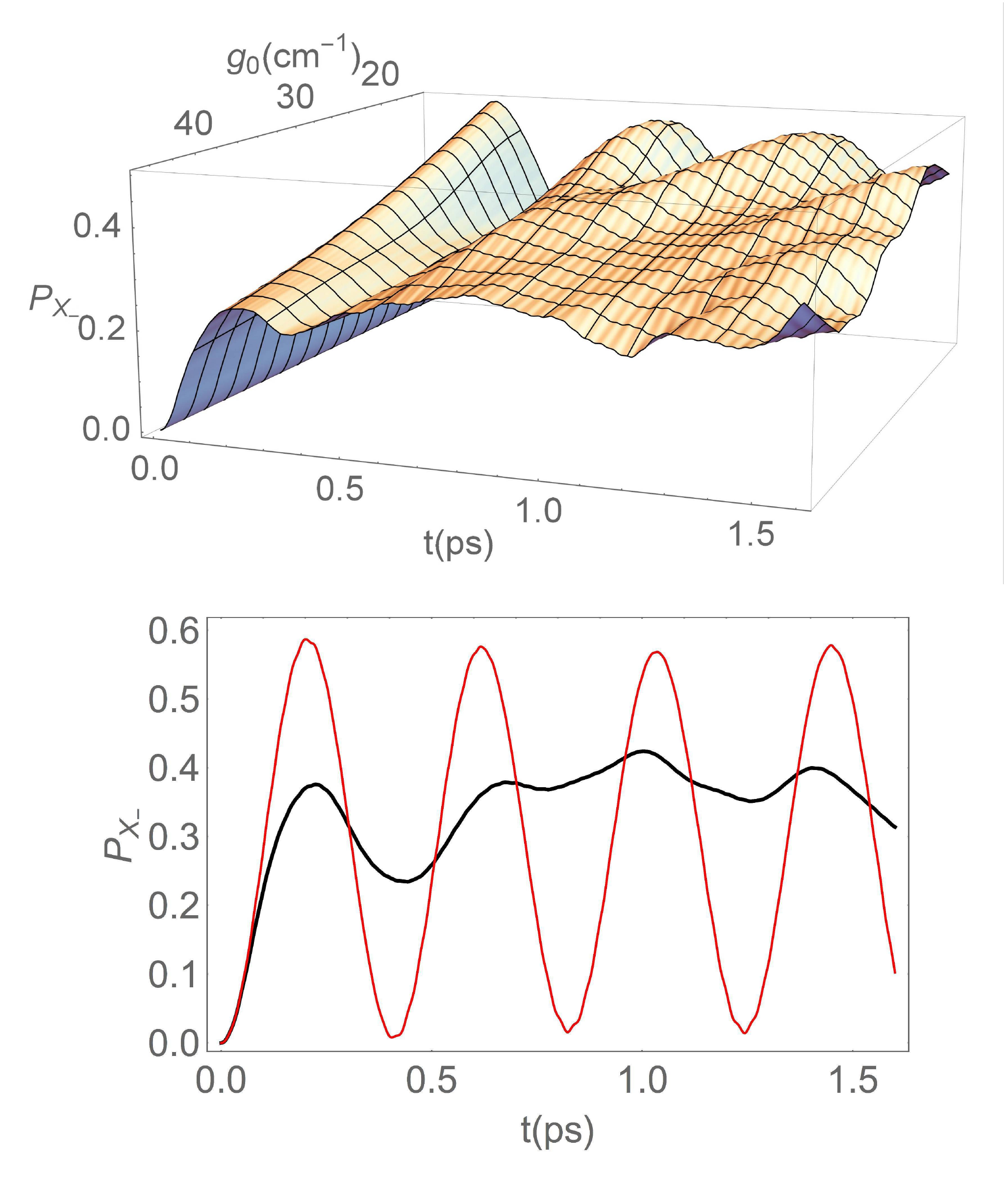}
\caption{Top panel: $P_{X_-}$ as a function of time and $g_0$. The system parameters are the ones given in Fig. \ref{fig1}, and $\omega_0=10^{-2} \omega_{{\rm vib}}$. Lower panel: $P_{X_-}$ as a function of time for $g_0=0$ (red) and for $g_0=10^{-1}g$ (black).}
\label{figyy}
\end{center}
\end{figure}

\begin{figure}
\begin{center}
\includegraphics[width=7cm]{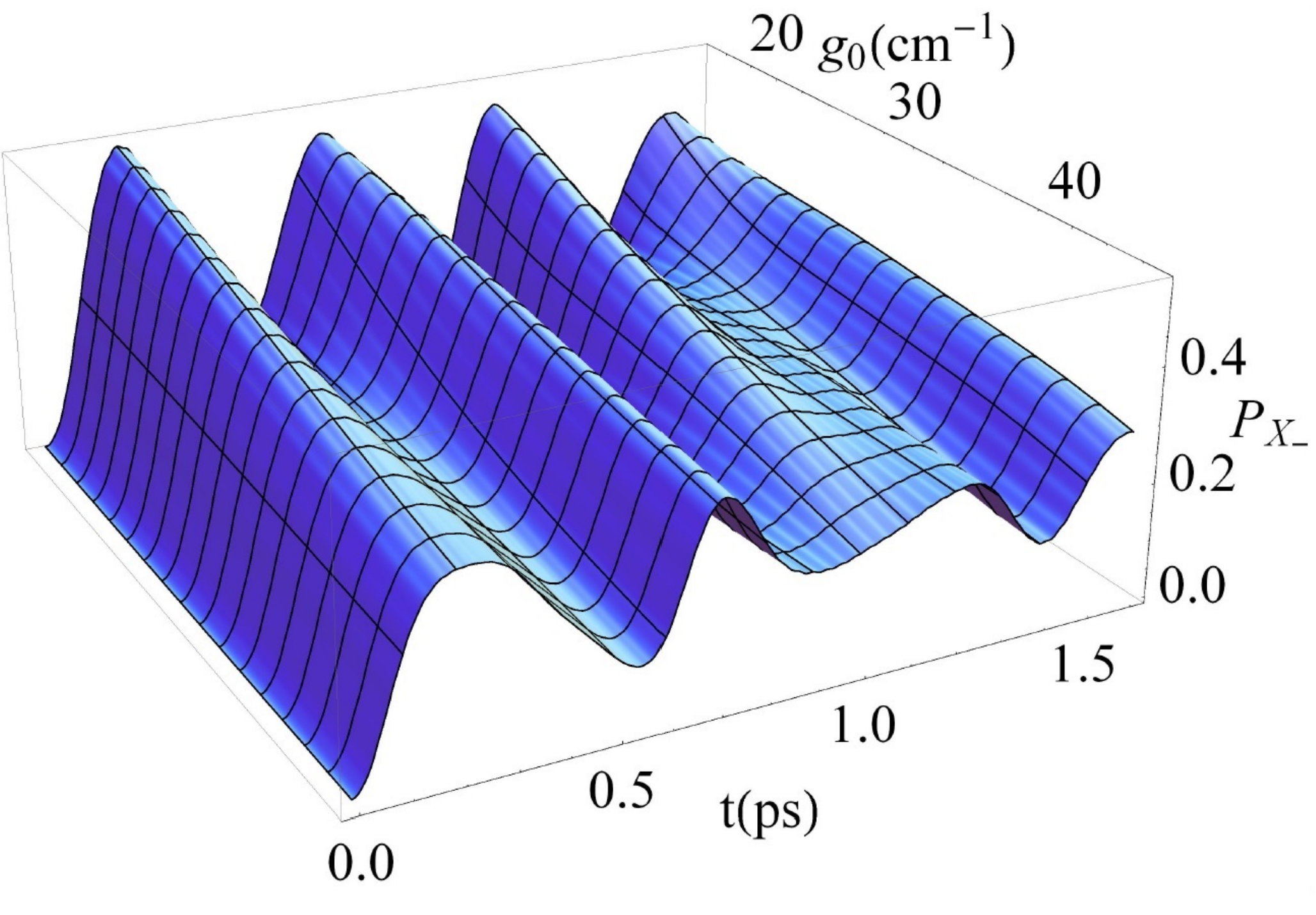}
\caption{ $P_{X_-}$ as a function of time and $g_0$ for $\omega_0=10^{-1} \omega_{{\rm vib}}$.}
\label{figyy2}
\end{center}
\end{figure}

Using the coupling $g_0$ and the frequency $\omega_0$ of $k_0$ as free parameters, we explored different regimes and found behaviours that are in qualitative agreement with the whole bath case. In Fig. \ref{figyy} we plot the population of the excitonic eigenstate $|X_- \rangle$, i.e. $P_{X_-}=\sum_n p_{n,X_-}(t)$ as a function of time and of $g_{0}$ for  $\omega_0=10^{-2}\omega_{{\rm vib}}$. In the range of  $g_0$ chosen (centered approximately around $g_0=10^{-1} g$), we observe the transition from the coherent regime to the incoherent one, where the spectral weight of the  $|X_+ ,0\rangle \to|X_-, 1\rangle$ is reduced. As already pointed out, within our simplified model,  $P_{X_-}$ does not reach a true stationary state. However, there is the clear tendency to get stabilized around a plateau. As expected,  if the frequency of that mode is too close to $\omega_{{\rm vib}}$, where the weak-coupling approximation breaks down, the single mode is not able to capture the essential features of the whole bath.
This is illustrated in Fig. \ref{figyy2}, where we have chosen $\omega_0=10^{-1}\omega_{{\rm vib}}$. In this regime, decoherence is expected to  take place before coherent oscillations are established \cite{oreilly14}, while we observe high-visibility oscillations.

\begin{figure}
\begin{center}
\includegraphics[width=7cm]{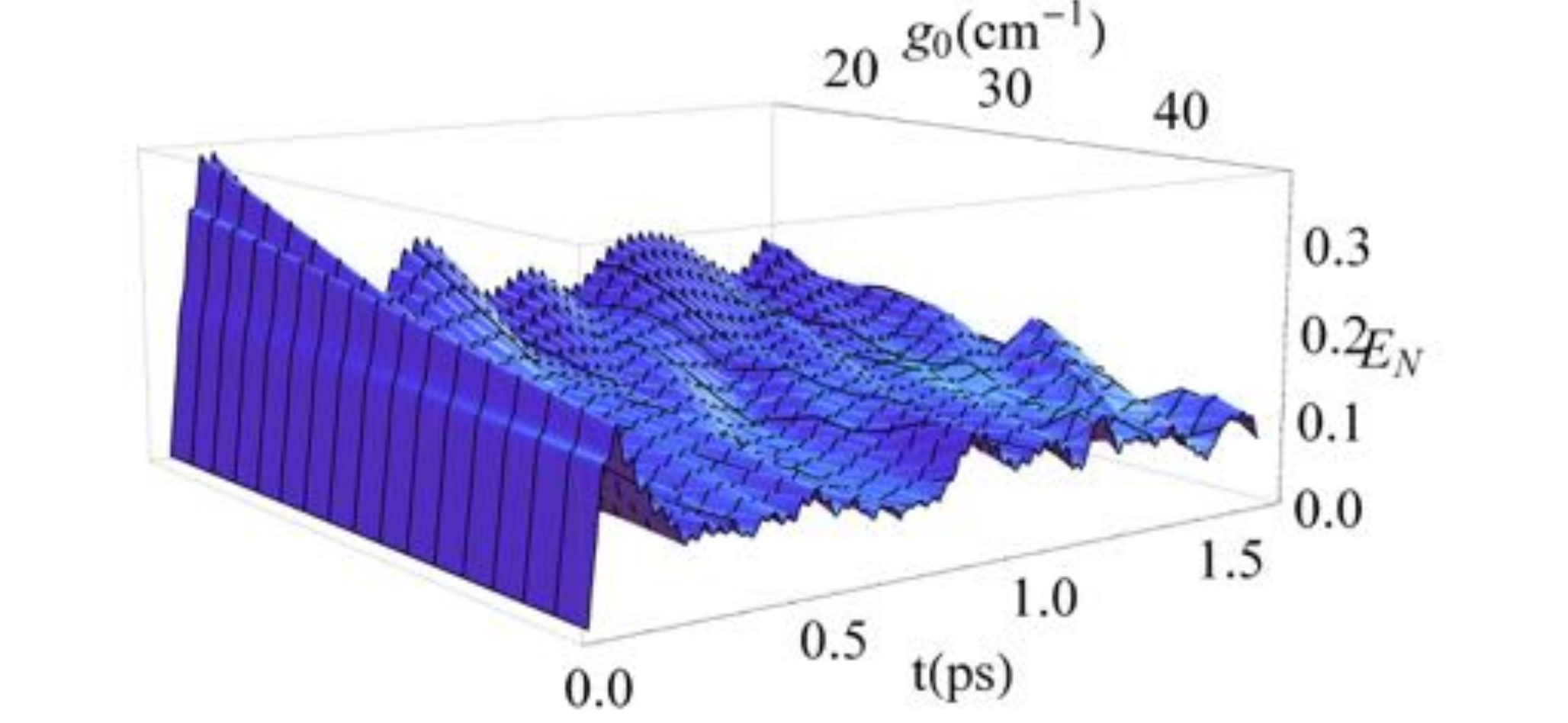}
\caption{Negativity of entanglement $E_N$ as a function of time and $g_0$ for $\omega_0=10^{-2} \omega_{{\rm vib}}$. }
\label{figent}
\end{center}
\end{figure}

\begin{figure}
\begin{center}
\includegraphics[width=7cm]{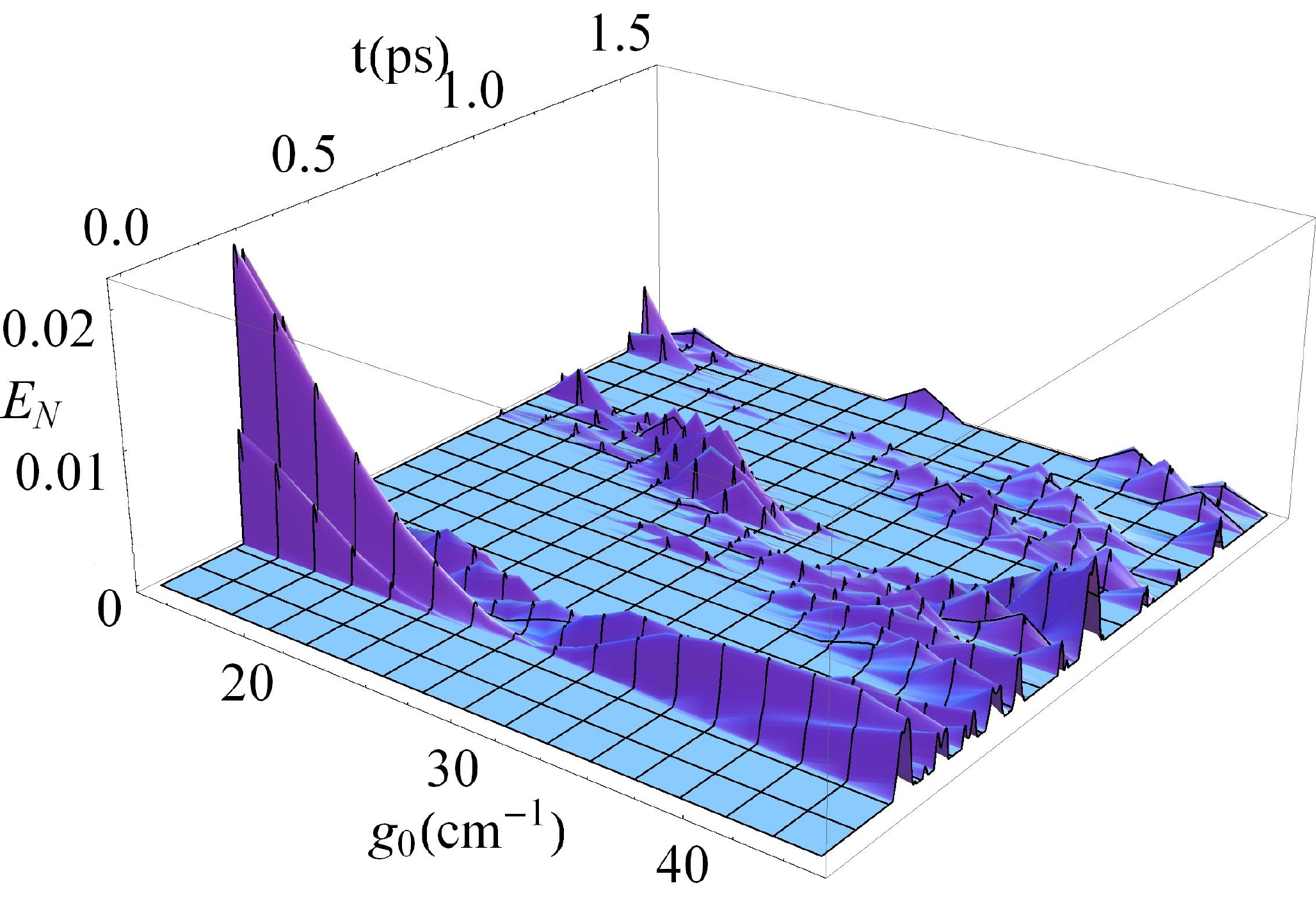}
\caption{Negativity of entanglement between the dimer and the low-frequency mode as a function of time and $g_0$ for $\omega_0=10^{-2} \omega_{{\rm vib}}$. }
\label{figent3}
\end{center}
\end{figure}

Once established the conditions under which modeling the bath as a single mode represents a suitable approximation, we use this technique to study the behaviour of entanglement negativity, represented in Fig. \ref{figent}, for the case where decoherence is expected not to completely suppress the coherent oscillation (Fig. \ref{fig1}). For the sake of clarity, here, we will omit the complementary discussion about entanglement of formation and discord, as there are no qualitative differences. As in the unperturbed case, we observe an initial growth after which the indicator starts going down. The important point is that $E_N$ remains positive at any time, showing the resilience of entanglement against noise. It may also be interesting to see whether the decrease of dimer-vibration entanglement is somewhat compensated   by the creation of entanglement between the dimer and the bath. Actually, apart from some special values of $g_0$ and $\omega_0$ and possibly because of the roughness of the single-mode approximation performed, this does not happen (see Fig. \ref{figent3}), or, at least, entanglement negativity does not reveal it. This means that a global loss of quantumness takes place even in the single-mode approximation.  Let us stress that, even if we have modelled the bath as a genuine quantum system, the results indicate that it actually behaves as a classical source of noise.   In fact, it induces decoherence in the exciton-vibration part of the model without getting quantum correlated itself.  Notice that we have considered a very small environment that unavoidably presents recurrences regimes and back-flow of information. The absence of bath-system quantum correlations is an interesting result, even if not completely unexpected  due to the fact that the  distribution  of entanglement among 
many parties is constrained by the property of monogamy and the major quantity of entanglement is already established between the dimer and the resonant mode \cite{ckw}.

\section{Conclusions}

Understanding the very fundamental mechanism responsible for high-efficiency energy transfer in photosynthesis is expected to lead to both fundamental and practical implications. On the one hand, excitation energy distribution is unavoidably influenced by the presence of environmental degrees of freedom, while, on the other hand, the role played by quantum mechanics in biological  structures has yet to be fully understood. 

We have investigated the presence of quantum correlations in a dimer-exciton system using negativity of entanglement and quantum discord. From a qualitative point of view, we have found traces of quantumness at any time of interest. This implies that quantum effects are even deeper than what indicated by the time behavior of the Mandel parameter or of the Glauber quasi-probability distribution.  Our  results suggest that the information conveyed by entanglement and  discord is richer and could be extended to different models and working regimes. It seems that the phase-space characterization of non-classicality is more sensitive to the main contribution to the dynamics, while entanglement and quantum discord also capture the multi-mode structure of the bosonic mode.
Let us stress that the phase-space parameters, as well as entanglement negativity (apart from some special case \cite{roncaglia}) are sufficient criteria for the existence of quantumness, while discord is both sufficient and necessary. Then, quantum correlations are established immediately after the interaction takes place. Let us point out that we have observed a correlation between transport and non-classical properties. This does not necessarily imply the existence of a causal relation, which we would not able to prove, and indeed such kind of answer  has not been found yet in all the existing literature of the field. Nevertheless, our results point out that quantum correlations can play a significant role and may stimulate further studies on the subject, eventually addressed to understand whether a functional role exists or not.

 It is also important to set the limits under which non-classicality is robust against thermal noise. We proposed a simplified approach to this problem based on the use of a single mode to mimic the role of the environment. The computational advantage of this method, compared to hierarchical expansions, is evident. As expected, this simplified approach is only meaningful in the the weak coupling limit. We tested the method in our system, found the regime where it can be applied, and used it to assess the quantum character of the dynamics also in the presence of noise. Unexpectedly, despite the high-degree of non-Markovianity and and back-flow of information induced by such a small environment, negligible quantum correlations are detected between the system and the bath itself. We have found a regime where  even a single mode acts as a classical source of noise as distribution of entanglement among many parties is subject to monogamy restraints.
 
Finally, let us stress that the very same model study can also be found in different  physical contexts. For instance, it could describe electron-phonon interaction in metals. It might also be interesting to analyze the dynamical behaviour by considering different  system parameters in order to establish whether the observed phenomenon is a general characteristic of the model or it requires specific experimental condition to be matched. 
 
%
%

\section*{Acknowledgements}
We are indebted to Mario Rasetti for useful discussions and comments. The  financial support of  Compagnia di San Paolo (Torino, Italy) in the frame of the INRIM project on \textquotedblleft Quantum Correlations"   is gratefully acknowledged.

\appendix
\section*{Appendix A}
The model introduced through Eqs. (\ref{HEL}-\ref{HELVIB})  is defined in the tensor product Hilbert space $\mathfrak{H}$ $=$ $\mathfrak{H}_1 \otimes \mathfrak{H}_2 \otimes \mathfrak{F}_1 \otimes \mathfrak{F}_2$, where $\mathfrak{H}_1$, $\mathfrak{H}_2$ are the two-dimensional Hilbert spaces of the chromophores and $\mathfrak{F}_1$, $\mathfrak{F}_2$ are the $\infty$-dimensional Fock spaces of the phonons. The analysis simplifies considerably by introducing the eigenstates $|X_\pm\rangle$ of $H_{\rm el}$,  $H_{\rm el} |X_\pm\rangle$ $=$ $\lambda_\pm |X_\pm\rangle$, with energy  splitting $\lambda_+ - \lambda_-$ $=$ $\sqrt{( \varepsilon_1-\varepsilon_2)^2+4V^2}$, and the collective phonon modes $b_{\pm}^{\dag}=(b^\dag_1 \pm b^\dag_2)/\sqrt{2}$, with $b_+^{\dag}$ ($b_-^{\dag}$) corresponding to the center-of-mass (relative displacement) phonon mode. In view of the properties of $\hat{N}$, the  effective chromophores Hilbert space reduces to the single two-dimensional one-particle (or spin $\frac{1}{2}$) space $\mathfrak{\tilde{H}}$, with $\mathfrak{H}_1 \otimes \mathfrak{H}_2$ $\longmapsto$ $\mathfrak{\tilde{H}}$. Furthermore, since the center-of-mass mode $b_{+}^{\dag}$ is not coupled to the electronic degrees of freedom, only the relative displacement bosonic operator $b_-^\dag$ is relevant to the system dynamics, so that the phonon space maps into the single $\infty$-dimensional Fock space $\mathfrak{\tilde{F}}$, i.e., $\mathfrak{F}_1 \otimes \mathfrak{F}_2$ $\longmapsto$ $\mathfrak{\tilde{F}}$. Eqs. (\ref{HEL}-\ref{HELVIB}) are combined resorting to the above simplifications and hence giving the effective exciton-vibration Hamiltonian introduced in Eq. (\ref{hsys}). There, the Pauli matrices $\sigma_z$ and $\sigma_x$ live in $\mathfrak{\tilde{H}}$.


\section*{Appendix B}
Given a bipartite density matrix $\varrho$ and its reduced states $\varrho_A={\rm Tr}_B\{\varrho \}$ and $\varrho_B={\rm Tr}_A\{\varrho \}$,  discord can be defined as the measure of how much
disturbance is introduced when trying to get information about party $A$ when party $B$ is measured \cite{discord}. It is defined as
\begin{equation}
{\cal D}_{A:B}={\cal I}(\varrho)-{\cal J}_{A:B},
\end{equation}
where ${\cal I}(\varrho)=S(\varrho_{A})-S(\varrho_{B})-S(\varrho)$ is the quantum  mutual information, obtained from its classical counterpart replacing the Shannon entropy with  the von Neumann entropy $S(\varrho)=-{\rm Tr}\{\varrho \log \varrho\}$,  and where the classical correlations  are given by
\begin{equation}
 {\cal J}_{A:B}= \max_{\{E_j^B\}}[S(\varrho_A)-S(A|\{E_j^B\})],\label{clas}
\end{equation}
 with the conditional entropy  $S(A|\{E_j^B\})=\sum_j p_j S(\varrho_{A|E_j^B})$, $p_j={\rm
Tr}_{AB}(E_j^B\varrho)$ and where $\varrho_{A|E_j^B}= E_j^B\varrho /{p_j} $ is the density
matrix after a POVM with elements $\{E_j^B\}$ has been performed on party $B$. Notice that the definition of discord is not symmetric under the exchange of the two parties. It is even possible to find states that are quantum-classical, that is, states that behave as quantum objects if one of the parties is observed and as classical objects by observing the other party.





\end{document}